\documentclass[pra,amsmath, amssymb, superscriptaddress, aps,twocolumn,letter]{revtex4-1}

\usepackage{color}
\usepackage{amsmath}
\usepackage{pifont}
\usepackage{amssymb}  
\usepackage{bbold}
\usepackage{float}
\usepackage{tikz}
\usepackage{makecell}
\usepackage{subfigure}
\usepackage{pifont}   
\usepackage{graphicx} 
\usepackage{dcolumn}  
\usepackage{bm}       
\usepackage{multirow} 
\usepackage{hyperref}
\usepackage{placeins}
\newcommand{\vect}[1]{\mathbf{#1}}

\newcommand{\comm}[2]{\left[{#1},{#2}\right]}

\newcommand{\norm}[1]{\left\|{#1}\right\|}

\newcommand{\yes}{\color{green}\ding{51}}%
\newcommand{\no}{\color{red}\ding{55}}%

\begin{document}

\renewcommand\floatpagefraction{0.8} 
\renewcommand\topfraction{0.8}       

\author{Michael Vogl}
\affiliation{Department of Physics, The University of Texas at Austin, Austin, TX 78712, USA}
\affiliation{Department of Physics, Northeastern University, Boston, MA 02115, USA}
\author{Pontus Laurell}
\affiliation{Department of Physics, The University of Texas at Austin, Austin, TX 78712, USA}
\affiliation{Center for Nanophase Materials Sciences, Oak Ridge National Laboratory, Oak Ridge, Tennessee 37831, USA}
\author{Aaron D. Barr}
\affiliation{Department of Physics, The University of Texas at Austin, Austin, TX 78712, USA}
\author{Gregory A. Fiete}
\affiliation{Department of Physics, The University of Texas at Austin, Austin, TX 78712, USA}
\affiliation{Department of Physics, Northeastern University, Boston, MA 02115, USA}
\affiliation{Department of Physics, Massachusetts Institute of Technology, Cambridge, MA 02139, USA}

\title{Analogue of Hamilton-Jacobi theory for the time-evolution operator}
\date{\today}

\begin{abstract}
In this paper we develop an analogue of Hamilton-Jacobi theory for the time-evolution operator of a quantum many-particle system. The theory offers a useful approach to develop approximations to the time-evolution operator, and also provides a unified framework and starting point for many well-known approximations to the time-evolution operator.  In the important special case of periodically driven systems at stroboscopic times, we find relatively simple equations for the coupling constants of the Floquet Hamiltonian, where a straightforward truncation of the couplings leads to a powerful class of approximations.  Using our theory, we construct a flow chart that illustrates the connection between various common approximations, which also highlights some missing connections and associated approximation schemes. These missing connections turn out to imply an analytically accessible approximation that is the ``inverse" of a rotating frame approximation and thus has a range of validity complementary to it. We numerically test the various methods on the one-dimensional Ising model to confirm the ranges of validity that one would expect from the approximations used. The theory provides a map of the relations between the growing number of approximations for the time-evolution operator.  We describe these relations in a table showing the limitations and advantages of many common approximations, as well as the new approximations introduced in this paper. 
\end{abstract}


\maketitle

\section{Introduction}
\label{sec:intro}
Our understanding of the dynamics of quantum many-particle systems and associated non-equilibrium phenomena has seen rapid growth in recent years \cite{RevModPhys.83.863,RevModPhys.89.011004}, which has resulted from advances in theory---especially Floquet systems \cite{Bukov2015,PhysRev.138.B979,10.1038/nphys4106,PhysRevLett.112.150401,PhysRevLett.118.260602,2018arXiv180500031W,PhysRevLett.116.250401,Ponte2015,PhysRevLett.114.140401,PhysRevLett.115.256803,PhysRevB.95.014112,PhysRevLett.116.120401,PhysRevX.7.011026}
---and experiment, particularly  in the preparation of and characterization of non-equilibrium states.\cite{RevModPhys.89.011004,RevModPhys.80.885,RevModPhys.83.1523,Basov2017,doi:10.1146/annurev-matsci-070813-113258,RevModPhys.83.471,doi:10.1080/00018732.2016.1194044,1402-4896-92-3-034004} As the field of non-equilibrium quantum systems expands, an increasing amount of effort is being devoted to the study of such systems, driven in part by the wealth of novel phenomenology that the time domain permits. For cold atoms trapped in optical lattices, time-dependent driving has enabled \cite{RevModPhys.89.011004,PhysRevA.91.033632} coherent control of tunneling \cite{PhysRevLett.100.040404}, induction of phase transitions \cite{PhysRevLett.102.100403,PhysRevLett.95.260404}, generation of effective magnetic fields \cite{PhysRevLett.107.255301}, and the measurement of nontrivial topological invariants \cite{2015NatPh..11..162A}. Noteworthy examples from solid state systems include photoinduced superconductivity, \cite{Fausti2011,doi:10.1080/00107514.2017.1406623} and hidden or otherwise inaccessible orders. \cite{Stojchevska2014, 2018arXiv180304490L,Gerasimenko2017} Even more strikingly, the time domain also allows entirely novel phases, such as time crystals,\cite{10.1038/nature21413,10.1038/nature21426} and non-equilibrium topological phases. \cite{PhysRevX.6.041001,PhysRevLett.116.250401,PhysRevB.79.081406,Lindner2011,PhysRevB.84.235108,PhysRevLett.107.216601,PhysRevX.3.031005,PhysRevB.90.115423} 


A number of numerical and analytical tools have been developed to understand the principal quantity of interest---the time evolution operator---which mathematically is a time-ordered exponential. It is not practical to summarize all known methods to calculate this quantity, so we will set our focus on analytically accessible approximations that can be used for arbitrary forms of time-dependence. This restriction will therefore exclude the vast literature on numerical methods, and a recently developed exact method \cite{Giscard2019} to calculate the time-evolution operator of finite-size systems.  Some approximations we can discuss in a unified manner with the framework introduced in this paper include the Dyson-Neumann series, \cite{PhysRev.75.486,PhysRev.75.1736,PhysRev.85.631} the Magnus expansion,\cite{Blanes2009,Feldm1984,CPA:CPA3160070404,0305-4470-22-14-019,0143-0807-26-1-015} Fer's series,\cite{0305-4470-22-14-019,623750276,0143-0807-26-1-015} Wilcox' series,\cite{0305-4470-22-14-019,doi:10.1063/1.1705306,0143-0807-26-1-015} the rotating frame approximation, \cite{PhysRevB.95.014112} and flow equation methods.\cite{vogl2018flow} 

The focus of this paper is on the flow equations for couplings (which we will introduce shortly) in a Hamiltonian and their relation to the various approximations mentioned above. We note that important results were obtained in prior work by flow equation methods in the equilibrium case, \cite{Wegner1994,Kehrein2007} 
and the non-equilibrium case \cite{PhysRevLett.111.175301} making use of the construct of a Sambe space.\cite{PhysRevA.7.2203} In this work, we find an analogue to Hamilton-Jacobi theory for the time-evolution operator that allows one to make clear connections between various approximation schemes. Our formulation takes shape in the form of flow equations for the couplings in the Hamiltonian.\cite{Wegner1994,Kehrein2007}

Our discussion culminates in a diagram, Fig.\ref{fig:diagram1}, that interrelates the different approximations and highlights spots which symmetry suggests can be filled by considering another limiting flow equation.  In this paper, we develop this limiting flow equation and find that the approximations perform as one would have expected from the order of approximations seen in the diagram in Fig.\ref{fig:diagram1}. We display a table, Table\ref{tableWhenTo}, that makes clear to the reader the range of validity for each approximation, along with its advantages and shortcomings. In particular, we find that the flow equations obtained in Ref.[\onlinecite{vogl2018flow}] are useful when truncated like Wegner's, \cite{Wegner1994,Kehrein2007} 
and the approximation we develop in this paper (by completing our flow chart) fills in a gap for an analytically accessible approximation in the intermediate time regime (or equivalently for a time-periodic system in the intermediate frequency regime).  Our results provide a comprehensive picture of the approximations to the time-evolution operator that can be used for time-periodic quantum many-particle systems. The approximation we develop here is particularly useful when $\norm{V}\ll \norm{H_0} \sim\omega $. That is, when the system is subject to a drive $V$ that is weak compared to the static part of the Hamiltonian $H_0$ but when the static part $H_0$ is not negligible when compared to the drive frequency $\omega$. This regime is relevant to e.g. cavity-QED applications \cite{Hafezi2013,PhysRevA.70.023819} (where strongly interacting photons are often subject to weak time-periodic drives), and weakly driven cold atom quantum ratchets.\cite{PhysRevA.82.023607}

Our paper is organized as follows. In Sec.\ref{sec:common_approx}, we give a short introduction to each of the common approximations mentioned above.  In Sec.\ref{sec:relations}, we establish relations between the different approximations.  In Sec.\ref{sec:reverse_rot} we present the new approximation and a diagram of relations in Fig.\ref{fig:diagram1} that were suggested by symmetry.  In Sec.\ref{sec:Ising} we compare the different approximations for an Ising model. In section \ref{sec:allApprox} we calculate the $l_2$ distance between the exact and the approximate time-evolution operators.    Finally, in Sec.\ref{sec:conclusions} we present our conclusions.  A few technical details are relegated to the appendices.

\section{Summary of common approximations to $U(t)$}
\label{sec:common_approx}
The time-evolution operator $U(t)$ fulfills
\begin{equation}
	i\partial_tU(t)=H(t)U(t); \quad U(0)=\mathbb{1}_{{H}},
	\label{timeevopeq1}
\end{equation}
where $\mathbb{1}_{{H}}$ is the identity on the Hilbert space of the Hamiltonian, ${H}$. We have set Planck's constant, $\hbar =1$. Eq.\eqref{timeevopeq1} can be solved by a simple matrix exponential if $[H(t_1),H(t_2)]=0$ for all $t_1$ and $t_2$. However, when $[H(t_1),H(t_2)]\neq 0$ Eq.\eqref{timeevopeq1}becomes more complicated and one has to concatenate matrix exponentials for infinitesimal time steps $dt$,
\begin{equation}
	U(t)=e^{-idtH(t-dt)}e^{-idtH(t-2dt)}\cdots  e^{-idtH(0)}.
	\label{trotterized}
\end{equation}
Such a procedure, which sometimes is called Trotterization, reproduces the formal solution: $U(t)={\cal T} e^{-i \int_0^t dt'H(t')}$ in the limit $dt\to 0$, where $\cal{T}$ is the time-order operator and the expression is called a time-ordered exponential.  However, in general this is not an analytically tractable procedure and therefore approximations are needed. In the remainder of this section we will summarize a few of the most common approximations.

\subsection{Dyson-Neumann series}
An important approximation to the time ordered exponential is due to Dyson.\cite{PhysRev.75.486,PhysRev.75.1736,PhysRev.85.631} One integrates Eq.\eqref{timeevopeq1} to find,
\begin{equation}
U(t)=\mathbb{1}_{H}-i\int_0^t dt' H(t')U(t').
\label{timeevopeq2}
\end{equation}
If one repeatedly reinserts the left side into the right side of the equation one obtains a series in powers of $H(t)$. Truncated at second order, this series is given as,
\begin{equation}
	U(t)=\mathbb{1}_{{H}}-i\int_0^t dt' H(t')-\int_0^t dt_1\int_0^{t_1}dt_2 H(t_1)H(t_2)+\mathcal{O}(H^3).
	\label{dysNeumann}
\end{equation}
Neglecting higher order terms in $H(t)$ is valid if $H(t)$ is small compared to $\sim 1/t$. In other words, such an expansion is restricted to sufficiently short times.  In addition, truncating the series destroys its unitarity, which is a serious drawback. The loss of unitarity occurs already at first order in $H(t)$.

If one is interested in the evolution of eigenstates of a constant Hamiltonian $H_0$, it is advantageous to split $U(t)=U_I(t)e^{-iH_0t}$ because in this case the second factor only leads to a phase $e^{-iH_0t}|\psi\rangle=e^{i\phi}|\psi\rangle$. For expectation values of an operator $\hat O$ the phase factor disappears $\langle \hat O\rangle(t)=\langle\psi|U^\dag(t)\hat OU(t)|\psi\rangle=\langle\psi|U_I^\dag(t)\hat OU_I(t)|\psi\rangle$ and $U_I$ encodes all relevant information. It now fulfills
\begin{equation}
 U_I(t)=\mathbb{1}_{{H}}-i\int_0^t dt' V_I(t')+\mathcal{O}(V_I^2),
\end{equation}
where $V_I=e^{iH_0t}(H(t)-H_0)e^{-iH_0t}$. Such a procedure is called the interaction picture (hence the ``$I$" subscript). The expansion for $U_I(t)$ is often used in setting up Feynman diagrams for scattering problems because for small perturbations $V(t)$ the procedure works well enough to approximate the S-matrix, $S=U_I(t=\infty)$.

\subsection{Magnus expansion}
The broken unitarity in Dyson's approach is a serious drawback because spurious terms may appear in calculations---a problem Magnus\cite{Blanes2009,Feldm1984,CPA:CPA3160070404,0305-4470-22-14-019,0143-0807-26-1-015} solved. His way of dealing with this issue was by making the ansatz $U(t)=e^{\Omega(t)}$ for the time-evolution operator and searching for anti-Hermitian $\Omega$ instead of $U(t)$. Inserting this ansatz into Eq.\eqref{timeevopeq1} and using the general expression for the derivative of the exponential map he found that,
\begin{equation}
	\frac{d\Omega(t)}{dt}=i\frac{ad_\Omega}{e^{ad_\Omega}-1}H(t),
\end{equation}
where  the shorthand $ad_\Omega=[\Omega,.]$ was used.

The solution can be found by first solving the equation for $ad_\Omega=0$, {\em i.e}. to lowest order and then reinserting the result to generate higher orders. One finds that,
\begin{equation}
\begin{aligned}
&U(t)= e^{\Omega_1(t)+\Omega_2(t)+\Omega_3(t)+\mathcal{O}(H^4)},\\
	&\Omega_1(t)=-i\int_0^t dt' H(t'),\\
	&\Omega_2(t)=-\frac{1}{2}\int_0^t dt_1\int_0^{t_1} dt_2[H(t_1),H(t_2)],\\
	&\Omega_3=\frac{i}{6} \int_0^t dt_1 \int_0^{t_1}d t_2 \int_0^{t_2} dt_3
	\Bigl(\left[H(t_1),\left[H(t_2),H(t_3)\right]\right]+(t_1\leftrightarrow t_3)\Bigr),
	\end{aligned}
	\label{Magnus}
\end{equation}
where $(t_1\leftrightarrow t_3)=\left[H(t_3),\left[H(t_2),H(t_1)\right]\right]$ is used as a shorthand. We denote higher orders by $\Omega_n$.

Similar to the case of the Dyson series, this approximation only works for sufficiently short times or sufficiently small Hamiltonians. However, it is an improvement in that it is unitary to all orders and therefore its mathematical structure is more sound. We note that at the lowest orders it agrees with the expansion discussed in Ref.[\onlinecite{PhysRevB.95.014112}].

It is worth noting that the Magnus Expansion does not converge in all cases, which means that an optimal cut-off order exists in these cases. Important work to understand how this affects the time-scales accessible by a description using a Magnus expansion has been published in
Ref.\cite{KUWAHARA201696}.

\subsection{Wilcox expansion}
Matrix exponentials of complicated operators are difficult to calculate and approximate. Wilcox (inspired by Fer's work, which we will discuss next) \cite{0305-4470-22-14-019,doi:10.1063/1.1705306,0143-0807-26-1-015} split the Magnus expansion into separate exponentials: $e^{\Omega_1(t)+...+\Omega_n(t)}\to e^{W_1(t)}...e^{W_n(t)}$, where $W_m=\mathcal O(H^m)$. Terms of order $e^{\mathcal{O}(H^n)}$ are neglected if the product is truncated after $e^{W_{n-1}}.$ Such an expansion would be advantageous if $e^{W_i}$ is easier to calculate than $e^{\sum_i \Omega_i}$.

To generate the terms up to $W_n$ one uses the Baker-Campbell-Hausdorf (BCH) formula $e^{A}e^{B}= e^{A+B+\frac{1}{2}[A,B]+\mathcal{O}(A^2B,AB^2)}$. One also introduces a dummy parameter $\delta$ that keeps track of different orders of $H$. One then finds that $e^{\delta W_1(t)}...e^{\delta^n W_n(t)}\approx e^{\sum_{m=1}^{n} \delta^m O_m(W_i)}$, where $O_m$ is a function of the operators $W_i$ that includes all terms of order $\delta^m$ that are generated by the BCH expansion. A comparison with $e^{\sum_m \delta^m \Omega_m}$ gives the set of equations $\Omega_m=O_m(W_i)$. These can be solved for the $W_i$. A more detailed description of this procedure is given in Ref. \cite{0305-4470-22-14-019}.

One finds to the first two orders,
\begin{equation}
\begin{aligned}
&U(t)=e^{W_1(t)}e^{W_2(t)}e^{\mathcal{O}(H^3)},\\
&W_1(t)=-i\int_0^t dt' H(t'),\\
&W_2(t)=-\frac{1}{2}\int_0^t dt_1\int_0^{t_1} dt_2[H(t_1),H(t_2)].
\end{aligned}
\label{Wilcox}
\end{equation}
These terms and higher order terms are related to the Magnus expansion. This relation up to order $H^3$ is given by
\begin{equation}
\begin{aligned}
	&W_1(t)=\Omega_1(t);\quad W_2=\Omega_2(t);\\ &W_3(t)=\Omega_3(t)-\frac{1}{2}[\Omega_1(t),\Omega_2(t)].
\end{aligned}
\end{equation}

We will see later in an explicit example illustrating that this approximation is not as good as the Magnus approximation. While we cannot be certain that this is generally the case, we have observed the same property in other cases. One can, however, make a good argument as to why this statement might be generally true. 

The behaviour may be rooted in the fact that the Wilcox approximation assumes that higher order terms $e^{W_n}$ appear to the right of lower order terms $e^{W_{n-1}}$ rather than the left. There is no {\em a priori} reason for either choice. This kind of asymmetry or ambiguity does not exist in the Magnus case, which may be the reason it performs better. A similar type of asymmetry that is due to an ambiguity of ordering also exists for the Zassenhaus formula $e^{t(X+Y)}= e^{tX}~ e^{tY} ~e^{-\frac{t^2}{2} [X,Y]}\cdots$, which is a dual of the BCH formula. Observations of this asymmetry have inspired recent attempts to symmetrize the Zassenhaus formula\cite{ARNAL201758}, which was able to yield some improvements in numerical accuracy. One may only wonder if a similar procedure could be used to improve on the Wilcox expansion. 

It is also worth noting that the procedure in Ref.\cite{ARNAL201758} introduces its own ambiguities. Rather than having to ask if one should order different terms in the perturbation series from left to right, or right to left, one has to ask if one should order them from inside to outside or from outside to inside. Therefore, the usefulness of the Wilcox approximation may be restricted to niche uses where separate operator exponentials $e^{W_i}$ are easier to compute than $e^{\sum_i \Omega_i}$.

\subsection{Fer's series}
Fer \cite{0305-4470-22-14-019,623750276,0143-0807-26-1-015} approached the challenge of finding a time-ordered exponential quite differently. His idea was to first ignore the time-ordering aspect and as a first approximation take,
\begin{equation}
	U(t)\approx U_1(t)=e^{-i\int_0^t dt' H(t')},   
\end{equation}
for a general time-dependent Hamiltonian $H(t)$.

Unlike Magnus he did not search for corrections to the exponent but instead took the time-evolution operator to factorize, $U(t)=U_1(t)U_2(t)$, and found an equation for $U_2(t)$,
\begin{equation}
	i\partial_t U_2(t)=H_2(t)U_2(t);\quad H_2(t)=U_1(t)^\dag (H(t)-i\partial_t)U_1(t). 
\end{equation}

One may now for $U_2$ again ignore the time-ordering aspect and repeat the same procedure. One then finds the recursive scheme,
\begin{equation}
\begin{aligned}
  &U(t)= U_1... U_n+\mathcal{O}(H^{2^n}),\\
  &U_j(t)=e^{-F_j(t)};\quad F_j(t)=i\int_0^t dt' H_{j}(t'),\\\
  &H_j(t)=U_{j-1}^\dag(t) (H_{j-1}(t)-i\partial_t)U_{j-1}(t);\quad H_1(t)=H(t).
\end{aligned}
\label{Ferscheme}
\end{equation}

The advantage of this procedure over the Wilcox approach is that at each order $j$ infinitely many orders of $H(t)$ are added to the exponents $F_j$. Including infinitely many orders of $H(t)$ should be expected to result in a more reliable approximation to the time-evolution operator. While infinitely many terms are added, the method is still controlled. It is found  \cite{0305-4470-22-14-019} that the terms that are neglected when dropping the $n$-th term in the product are of order $\mathcal{O}(H^{2^n})$. For these reasons (in our example later) the Fer expansion is found to be more reliable than the Magnus expansion if we break the series off at small orders, which is the practical thing to do because high orders become complicated in both cases. Therefore, while the convergence radius of the Fer approximation is smaller than for the Magnus case \cite{Blanes2009} at small orders, even when it does not converge, it is often found to be more reliable---an effect often-times humorously(!) summarized by Carrier's rule: ``Divergent series converge faster than convergent series because they don't have to converge.'' \cite{Boyd1999} 

The disadvantage of the method over the Magnus case is that it is often extremely difficult to calculate the $H_j$, in many cases even $H_2$. Furthermore, the method---like the other approaches discussed to this point---is restricted to relatively short times.

\subsection{Rotating frame approximation}
\label{Rotframesection}
In the rotating frame approximation, one finds an approximate time-evolution operator without having to worry about the time-ordering aspect.  This is accomplished by removing the time-dependence up to an \emph{arbitrarily} chosen time $T$ by a unitary transformation. In general this is a difficult task. However, it turns out that it is possible to do this to a good approximation if one splits $H(t)=\bar H_T+V_T(t)$ into a part that is constant on the interval $[0,T]$ given by $\bar H_T=\frac{1}{T}\int_0^{T} dt H(t)$, and a part $V_T(t)=H(t)-\bar H_T$ that averages to zero over the same interval. Here, $T$ is an arbitrarily chosen time at which the time evolution operator will be evaluated. The time dependence can be arbitrary and there is no need for a time-periodic drive.

However, it is important to note that this type of splitting occurs naturally in periodically driven systems at stroboscopic times. For this reason this type of approximation is commonly applied in a Floquet setting. One should also note that statements about the convergence of stroboscopic time evolution operators in the literature apply for a non-periodic case if we take $T$ as an ``artificial'' stroboscopic time and ensure to exclude all statements that consider multiple periods. This point is illustrated by realizing that, to calculate $U(T)$, we do not need to know if $H(2T)=H(T)$---this does not have to be true, rather it is enough to know only about times $0<t<T$.

To remove the time dependent part of the Hamiltonian one may apply the unitary transformation,
\begin{equation}
	S_{1,T}(t)=e^{-i\int_0^t dt' V_T(t')},
\end{equation}
to Eq.\eqref{timeevopeq1}, which achieves the goal of removing $V(t)$ if the time interval $[0,T]$ is comparatively small. A particularly convenient property of this transformation is that it reduces to the identity operator at $t=T$ by construction: $S_{1,T}(T)=\mathbb{1}_{{H}}$. 

Because of this property the time-evolution operator $U$ at times $T$ has no contribution from $S_{1,T}(t)$. Therefore, to learn about times $T$ it is enough to calculate the time evolution operator $U_T$ in the rotating frame,
\begin{eqnarray}
	i\partial_tU_T(t)&&=H_{1,T}(t)U_T(t),\\
	 \quad H_{1,T}(t)&&=S_{1,T}^\dag (H(t)-i\partial_t) S_{1,T}.
\end{eqnarray}

A solution at times $T$ that ignores the time-ordering aspects of the time evolution operator,
\begin{equation}
	U(T)\approx e^{-i\int_0^{T} dt H_{1,T}(t)}
\end{equation}
 in many cases now turns out to be an improvement over the same done for Eq.\eqref{timeevopeq1}---particularly this is true if $V_T(t)$ is larger than $\bar H_T$ \cite{PhysRevB.95.014112,vogl2018flow}.

As with Fer's method, one may iterate this procedure. One may split $H_{1,T}(t)$ in the same way we split $H(t)$. Following this logic, one finds that the time evolution operator can be successively approximated. An iterative procedure is given by,
\begin{equation}
\begin{aligned}
&U(t)\approx \left.e^{-i\int_0^{T} dt H_{n,T}(t)}\right|_{T=t},\\
&H_{n,T}(t)=S_{n,T}^\dag(t) (H_{n-1,T}(t)-i\partial_t) S_{n,T}(t),\\
&S_{n,T}=e^{-i\int dt V_{n,T}(t)},\\
& V_{n,T}(t)=H_{n,T}(t)-\frac{1}{T}\int_0^{T} dt H_{n,T}(t);\quad H_{0,T}(t)=H(t).
\end{aligned}
\label{fullrotframe}
\end{equation}

One finds \cite{PhysRevB.95.014112,vogl2018flow} that this approximation offers a significant improvement over the Magnus approximation when $V(t)$ is large. However, it is sometimes more cumbersome to implement.  Also, it is important to stress that, since $T$ could be chosen arbitrarily, as a final step one has to set $T=t$ in $U(t)$.

\section{Relations between the methods}
\label{sec:relations}
While the approximations discussed in Sec.\ref{sec:common_approx} may seem unrelated to one another at a first glance, there is an overarching reformulation of Eq.\eqref{timeevopeq1} that connects them all.

Let us recall the basic idea of Hamilton-Jacobi theory \cite{goldstein2002classical}. In Hamilton-Jacobi theory one arrives at a reformulation of classical mechanics by searching for a generating function of canonical transformations that make the (generally time-dependent) Hamiltonian equal to a constant---thereby removing the focus from Hamilton's equations. The full information of the dynamics is then absorbed into the generator. Without loss of generality, the constant may be taken to be zero.  Specifically, one may consider a generator $S(\vect q,\vect P,t)$ of canonical transformations from coordinates $(\vect q,\vect p)$ to coordinates  $(\vect Q,\vect P)$.  One finds that the Hamiltonian $K(\vect Q,\vect P,t)$ in terms of the new coordinates is given as
\begin{equation}
	K(\mathbf{Q},\mathbf{P},t) = H(\mathbf{q},\mathbf{p},t) + {\partial S \over \partial t}. 
	\label{CanTransf_newHam}
\end{equation}

The variables then also fulfill the conditions,
\begin{equation}
	\mathbf{p} = {\partial S \over \partial \mathbf{q}}; \quad
	\mathbf{Q} = {\partial S \over \partial \mathbf{P}}.
\end{equation}
One may then construct an $S$ such that $K=0$, and therefore $\dot {\vect P}=\dot {\vect Q}=0$, by solving
\begin{equation}
	H(\mathbf{q},{\partial S \over \partial \mathbf{q}},t) + {\partial S \over \partial t}=0,
	\label{Ham-Jac_equ_Cl}
\end{equation} 
where we made use of  $\mathbf{p} = {\partial S \over \partial \mathbf{q}}$.

One should stress that the key idea of the Hamilton-Jacobi formalism is to construct a coordinate transformation that gets rid of the Hamiltonian and thereby makes the equations of motion trivial.  We will do the same here. We will try to find a unitary transformation that gets rid of the Hamiltonian so that the equation for the time-evolution operator becomes trivial.  This is the central idea of this paper.

We take Eq.\eqref{timeevopeq1} as a starting point and introduce a unitary transformation, $S=e^{\delta s\Sigma(t)}$, generated by an as yet undetermined quantity $\Sigma(t)$ that will be chosen to reduce the Hamiltonian $H(t)$ as $H(t)\to (1-\delta s)H(t)$.  Hereby $\delta s$ is infinitesimal and ensures that the exponential can be safely expanded to lowest order. 

We may split the time evolution operator as $U_0=S^\dag U_{\delta s}=[1-\delta s\Sigma(t)]U_{\delta s}$ and act with $S(t)^\dag=1-\delta s\Sigma(t)$ from the left on the Schr\"odinger equation. The time evolution operator in the new frame $U_{\delta s}$ now fulfills the modified equation,
\begin{equation}
i\partial_tU_{\delta s}= \left (H(t)-i\delta s\partial_t \Sigma(t)-\delta s\comm{\Sigma(t)}{H(t)}\right)U_{\delta s}.
\end{equation}
One may read off a new Hamiltonian,
\begin{equation}
H(t,\delta s)=H(t)-i\delta s\partial_t \Sigma(t)- \delta s\comm{\Sigma(t)}{H(t)},
\label{eqHam}
\end{equation}
where we introduced a second parameter slot for a parameter $s$ in addition to the time dependence, which labels the behavior of the Hamiltonian along a unitary flow. If we replace $H(t)\to H(t,s)$, $H(t,\delta s)\to H(t,s+\delta s)$ and $\Sigma(t)\to \Sigma(t,s)$ we may keep track of how the Hamiltonian changes under a chain of dynamically determined infinitesimal unitary transformations, 
\begin{equation}
H(t,s+\delta s)=H(t,s)-i\delta s\partial_t \Sigma(t,s)- \delta s\comm{\Sigma(t,s)}{H(t,s)}.
\label{eqHamb}
\end{equation}
By a Taylor expansion around $\delta s=0$ we see that the Hamiltonian fulfills the differential equation,
\begin{equation}
\frac{dH(t,s)}{ds}=-i\partial_t \Sigma(t,s)- \comm{\Sigma(t,s)}{H(t,s)},
\label{eqHamc}
\end{equation}
which is the quantum analogue of Eq.\eqref{CanTransf_newHam} since both equations determine a transformed Hamiltonian. 
Unlike the classical version, we introduced an additional parameter $s$ for calculational convenience. The reason is that transformations in the quantum case are operators rather than phase space functions, and thus harder to determine. In principle, it could be possible to find the unitary transformation removing $H(t)$ in a single step, which would obviate the need to introduce $s$.

The appropriate boundary conditions are set by putting $H(t,0)$ as the original untransformed Hamiltonian. We may also keep track of the time-evolution operator in the original frame. For the first infinitesimal step it is,
\begin{equation}
	U(t)=S(t,\delta s) U_{\delta s}(t),
	\label{eq:S_U}
\end{equation}
and the more general case is found by repeating this after each infinitesimal transformation.

Up to this point, the treatment coincides with the use of time-dependent generators in Ref.[\onlinecite{0295-5075-93-4-47011}].  We now, however, choose $\Sigma$ very different from the Wegner generator (which is designed to block diagonalize $H$). We choose it such that it reduces the Hamiltonian $H(t)\to (1-\delta s)H(t)$ by some infinitesimal value $\delta s$, 
\begin{equation}
\Sigma(t)=-i\int_0^t dt' H(s,t').
\label{eq:generator}
\end{equation}
Notice that this generator also leaves a residual term $\delta s \int_0^t dt_1\comm{ H(t_1)}{H(t)}$ in Eq.\eqref{eqHam}.  We will discuss it later. 

With our specific choice of generator $\Sigma$ we find that
Eq.\eqref{eqHamc} becomes,
\begin{equation}
\frac{dH(s,t)}{ds}=-H(s,t)+i\int_0^t dt_1\comm{H(s,t_1)}{H(s,t)},
\label{exact_flow}
\end{equation}
which is the equivalent of Eq.\eqref{Ham-Jac_equ_Cl} in the Hamilton-Jacobi formalism, just for the time evolution operator. The analogy becomes more clear if we recognize that both equations determine a transformed Hamiltonian that is zero. 

Let us discuss this in slightly more detail. One can directly check that the fixed point $H(s,t)=0$, which we want to reach, is stable. To see this one realizes that near the fixed point Eq.\eqref{exact_flow} reduces to $\frac{dH(s,t)}{ds}=-H(s,t)$. This means that near the fixed point $[\frac{dH(s,t)}{ds},H(s,t)]=0$. Thus the quantities display a behaviour like a scalar. Therefore, one may apply an ordinary fixed point analysis, according to which the fixed point $H(s,t)=0$ is stable. Furthermore, this fixed point is the only fixed point. Let us briefly see why. 

The equation that determines the fixed point is $H(s,t)=i\int_0^t dt_1\comm{H(s,t_1)}{H(s,t)}$. One may also realize that $\mathrm{Tr}(H(s,t)\comm{H(s,t_1)}{H(s,t)})$ is zero by the cyclic property of trace, and therefore the Frobenius inner product $\langle H(s,t),\comm{H(s,t_1)}{H(s,t)}\rangle=0$. Both sides of the equation, therefore, are perpendicular in an operator product sense. This means the equation can only be fulfilled if both sides are zero. Therefore, the only fixed point is $H=0$. Thus, it can be expected that the equations will flow toward $H(s,t)=0$ as needed for the analogy to the Hamilton-Jacobi theory to hold.

One should note that the unitary transformation that gets rid of the Hamiltonian was obtained by multiplying infinitely many  infinitesimal unitary transformations  $s_\delta(t,s)=(\mathbb{1}-i\delta s\int_0^t dt' H(s,t'))$. In other words, one may write,
\begin{equation}
\begin{aligned}
	S(t,s+\delta s)&=s_\delta(t,s) S(t,s)\\
	&=S(t,s)-i\delta s \int_0^t dt^\prime H(s,t^\prime)S(t,s).
\end{aligned}
\end{equation}
One finds via a Taylor expansion that,
\begin{equation}
i\partial_sS(t,s)=\int_0^t dt'H(s,t')S(t,s).
\label{transform}
\end{equation}
The time-evolution operator in the frame after rotation by $S$ is now trivially given as $U_{s=\infty}=\mathbb{1}$, because the Hamiltonian is zero. Therefore, the time evolution operator in the original frame is just $U(t)=S(t)$, by Eq.\eqref{eq:S_U}.

How do we make practical use of the operator valued Eq.~\eqref{eqHamc}? In it, $H(s,t)$ is a linear operator and therefore may be written as a linear combination of operators with coefficients $c_i(s,t)$, $H(s,t)=\sum_i c_i(s,t)\hat O_i$. This mathemetical structure in turn also implies that $-H(s,t)+i\int_0^t dt_1\comm{H(s,t_1)}{H(s,t)}=-\sum_i g_i(t,[c_j(s,t^\prime)])\hat O_i$,  where $g_i$ has a functional dependence on the $c_j(s,t')$ because $V(s,t)$ itself depends on the $c_j(s,t)$ and it appears under an integral.  

One may therefore write Eq.~\eqref{exact_flow} as,
\begin{equation}
\frac{dc_i(s,t)}{ds}=-\sum_i g_i(t,[c_j(s,t^\prime)]).
\label{coupling_eq}
\end{equation}
At a first glance one may wonder whether Eq.\eqref{exact_flow} is useful because it is a complicated functional equation making it difficult to obtain $H(s,t)$. Moreover, it may appear that not only did we {\em not} get rid of the problem of having to find a time ordered exponential Eq.\eqref{transform}, but we made the issue worse by adding additional complications.

However, this exercise was a worthwhile time investment because it allows a very simple way of identifying approximations. If we assume that we can get rid of $H(t)$ swift enough that $s\approx 0$ we may set $H(s,t)\approx H(0,t)=H(t)$ in the generator $\Sigma$ in Eq.\eqref{eq:generator}. Details on this kind of approximation are given in Ref.[\onlinecite{vogl2018flow}]. This means that we are left with
\begin{equation}
	\frac{dH(s,t)}{ds}=-H(t)+i\int_0^t dt_1\comm{H(t_1)}{H(s,t)},
	\label{flowingFer}
\end{equation}
where the complication of functional dependences on $t$ is gone. Now, if we let $s$ run from $0$ to $1$ we get rid of $H(t)$ to lowest order. The value of $s=1$ at this point may seem arbitrarily chosen. The heuristic reason, however, is that setting $s=0$ in the generator already assumes a small value of $s$. This is only justified if $s<1$, so for consistency we need to set $s<1$. The reason we let it run up to this maximum value is that we want to be able to get as close to a fixed point as possible. For a more detailed discussion of such an approximation we point to Ref.\cite{vogl2018flow}.  Another way to look at this approximation is to recognize that it just performs the unitary transform $S_F^1=e^{-i\int_0^t dt H(t)}$. This is the same thing that happens in the lowest order of Fer's approximation.

Since in a unitary transformation we do not lose any information one can make repeated use of Eq.\eqref{flowingFer}. Concatenating these transformations allows us to reconstruct the expansion due to Fer, Eq.\eqref{Ferscheme}. In fact, this reformulation is more powerful than the standard approach due to Fer since his method usually cannot be implemented analytically. The necessary unitary transformations are often hard to calculate. The advantage of our method is that we may make use of the non-perturbative nature of Fer's approach but avoid some of its difficulties if we make another non-perturbative approximation, which is taking a truncated ansatz for $H(s,t)$. This allows us to do the necessary unitary transform approximately while keeping infinite orders from the couplings in $H(t)$. The validity of such an approach will be shown later on in an explicit example.

The lowest order Wilcox approximation, $U(t)\approx e^{W_1}e^{W2}$, also follows naturally from Eq.\eqref{flowingFer}. If one solves Eq.\eqref{flowingFer} while neglecting the $H^2$ term one finds,
\begin{equation}
	H(s,t)\approx (1-s)H(t).
\end{equation}

Reinserting this result in Eq.\eqref{flowingFer}, one finds the solution
\begin{equation}
	H(1,t)=-\frac{i}{2}\int_0^tdt_1[H(t_1),H(t_2)].
\end{equation}

Therefore, to order $H$ the time evolution operator is given by the unitary transformation that we tried to implement, $U(t)=e^{-i\int dt' H(t')}$, which means that we reproduced the exponent $W_1$. Removing $H(1,t)$ by the same procedure, we reproduce $W_2$. Therefore, finding the lowest order Wilcox approximation from Eq.\eqref{exact_flow} was just as easy.  One should remark that we may also reproduce the Dyson-Neumann approximation, which is found by a Taylor expansion.

So how do we reproduce the remaining approximations - rotating frame and Magnus approximation by going back to Eq.\eqref{exact_flow}? Let us, like in the rotating frame approximation in Sec.\ref{Rotframesection}, split $H(s,t)=\bar H_T(s)+ V_T(s,t)$ into a constant part $\bar H_T(s)=\frac{1}{T}\int_0^T dt H(s,t)$ and a time-dependent part $V_T(s,t)=H(s,t)-\bar H_T(s)$ that averages to zero on the interval $[0,T]$. Any arbitrary choice of $T$ is possible; in a last step one will have to set $T=t$. If we assume that $V(t)$ is dominant in the generator Eq.\eqref{eq:generator} then Eq.\eqref{exact_flow} reduces to
\begin{equation}
\frac{dH(s,t)}{ds}=-V_T(s,t)+i\int_0^t dt_1\comm{V_T(s,t_1)}{H(s,t)},
\label{exact_flow2}
\end{equation}
where we stress that $s\in [0,\infty)$.

One should note that Eq.\eqref{exact_flow2} is not an approximation but rather a unitary transformation that achieves a different goal than the one previously considered. For the specific case of a periodically driven system we discussed it great detail in Ref.[\onlinecite{vogl2018flow}]. However, let us quickly summarize. This equation does not remove $\bar H(s)$  but only $V_T(s,t)$ will be removed. The generator $\Sigma=-i\int_0^T dt V_T(s,t) $ has an advantage over the original generator because it vanishes at times $T$. This makes the equation a bit more useful than Eq.\eqref{exact_flow} because the time evolution operator $U(t)$ at times $t=T$ now coincides with the time evolution operator  $U_{s=\infty}(T)$ in the rotated frame. It simply becomes,
\begin{equation}
	U(t)=\left.U_{s=\infty}(T)\right|_{T=t}=\left.e^{-iH(\infty,T)T}\right|_{T=t}.
\end{equation}

Since $T$ could be chosen arbitrarily this poses no restriction and we were able to set $T=t$ in a last step. With Eq.\eqref{exact_flow2} one may find the generator $H(\infty,t)$ of the time evolution.

Let us pause for a moment and realize that this choice of unitary transformation completely removed the need to calculate  a time-ordered exponential. One now just has to calculate a mundane matrix exponential. However, interpreting Eq.\eqref{exact_flow2} in the same way as an equation of the form Eq.\eqref{coupling_eq} did for the couplings $c_i$, we have traded the complications of a time-ordered exponential for flows of couplings with a complicated functional dependence.

Nevertheless, even in the functional form, Eq.\eqref{coupling_eq}, is useful when describing many-body driven systems because one can make an ansatz for $H(s,t)$ and one only has to solve a finite set of equations numerically for the couplings in $H(s,t)$. Semi-analytic calculations with such an expression for the time-evolution operator are then possible because a matrix exponential is much more accessible than a time ordered exponential. The method is particularly useful when dealing with periodically driven systems because $H(s=\infty)$ is then the Floquet Hamiltonian.

We should also note that  Eq.\eqref{exact_flow2} simplifies significantly for a specific class of Floquet systems. The functional dependence on couplings in Eq.\eqref{coupling_eq} vanishes completely for such a special case. Indeed if $H(t)$ is monochromatically driven, {\em i.e.} has the form $H(t)=H_0+e^{i\omega t}H_{+}+e^{-i\omega t}H_{-}$ with $H_+=H_-^\dag$, then for stroboscopic times $T=2\pi/\omega$ (where $\omega$ is the drive frequency) there are no functional dependences. Rather, by a comparison of coefficients one finds the set of equations
\begin{equation}
	\begin{aligned}
		&\frac{dH_{0}}{ds}=\frac{2}{\omega}[H_+,H_-]+\frac{1}{\omega}[H_0,H_+-H_-],\\
		&\frac{dH_{\pm }}{ds}=-H_{\pm}\pm\frac{1}{\omega}[H_{\pm},H_0-H_{\mp}].
	\end{aligned}
	\label{exactflowexample}
\end{equation}

We would like to stress the added convenience this result is expected to provide for numerical studies with the flow equation approach. One can now solve a set of differential equations for couplings of the generator of stroboscopic time evolutions, {\em i.e.} the Floquet Hamiltonian.

Now let us show the usefulness this approach provides when we try to find approximation schemes. We can make the same approximation to Eq.\eqref{exact_flow2} as we previously did to Eq.\eqref{exact_flow}.  Namely, we assume that we can get rid of $V(t)$ swiftly enough that we may set $V(s,t)\approx V(0,t)=V(t)$ in the generator. In this case Eq.\eqref{exact_flow2} simplifies to 
\begin{eqnarray}
	\frac{dH(s,t)}{ds}=-V_T(t)+i\int_0^t dt_1\comm{V_T(t_1)}{H(s,t)}.
	\label{Rotatingframeflow}
\end{eqnarray}
Again, to lowest order, $V(t)$ is removed if we let $s$ run from zero to one. This implements the unitary transformation $S_T=e^{-i\int dt V_T(t)}$, which vanishes at times $T$. Therefore, the time evolution operator in this approximation at times $t$ is given as,
\begin{eqnarray}
	U(t)&&\approx e^{-i\int dt' H_1(t')};\\
	\quad H_1(t)&&=S_T^{\dag}(t)(H(t)-i\partial_t)S_T(t)|_{T=t},
\end{eqnarray}
which is the same as the rotating frame approximation, and where we set $T=t$ because $T$ could be chosen arbitrarily to match $t$. Repeatedly applying the flow equations produces the full expansion Eq.\eqref{fullrotframe}. 

One should stress again that the advantage of the flow equation approach is that one may make a truncated ansatz for $H(s,t)$ and therefore calculate an approximate rotating frame approximation in cases where an exact matrix exponential may not be calculated. That is, we may take advantage of the non-perturbative nature of the rotating frame transformation in more cases. In Ref.[\onlinecite{vogl2018flow}] such a truncation for one model was discussed and one may see the advantage this approach still has over a Magnus expansion. We will see this explicitly for an example problem later in this work.

Now let us see how the lowest orders of the Magnus approximation can be obtained from this approach. As in the Wilcox approximation case, we solve Eq.\eqref{Rotatingframeflow} while dropping the commutator term to find,
\begin{equation}
	H(s,t)\approx \bar H+(1-s)V(t).
\end{equation}

If we reinsert this into Eq.\eqref{Rotatingframeflow} and perform an integration by parts we find that
\begin{equation}
	H(1,T)=\frac{1}{T}\int_0^{T}dt H(t)+\frac{1}{2T}\int_0^{T}dt\int_0^{t}dt_1[H(t),H(t_1)].
\end{equation}
The matrix exponential $U(T)\approx e^{-iH(1,T)T}$ is then the second order result of the Magnus expansion.  Lastly, the Dyson series to the low orders can be found, similar to the Wilcox approximation, by expanding $U(T)$ to order $H^2$. 

\begin{figure*}
	\centering
	\includegraphics[width=1.0\textwidth]{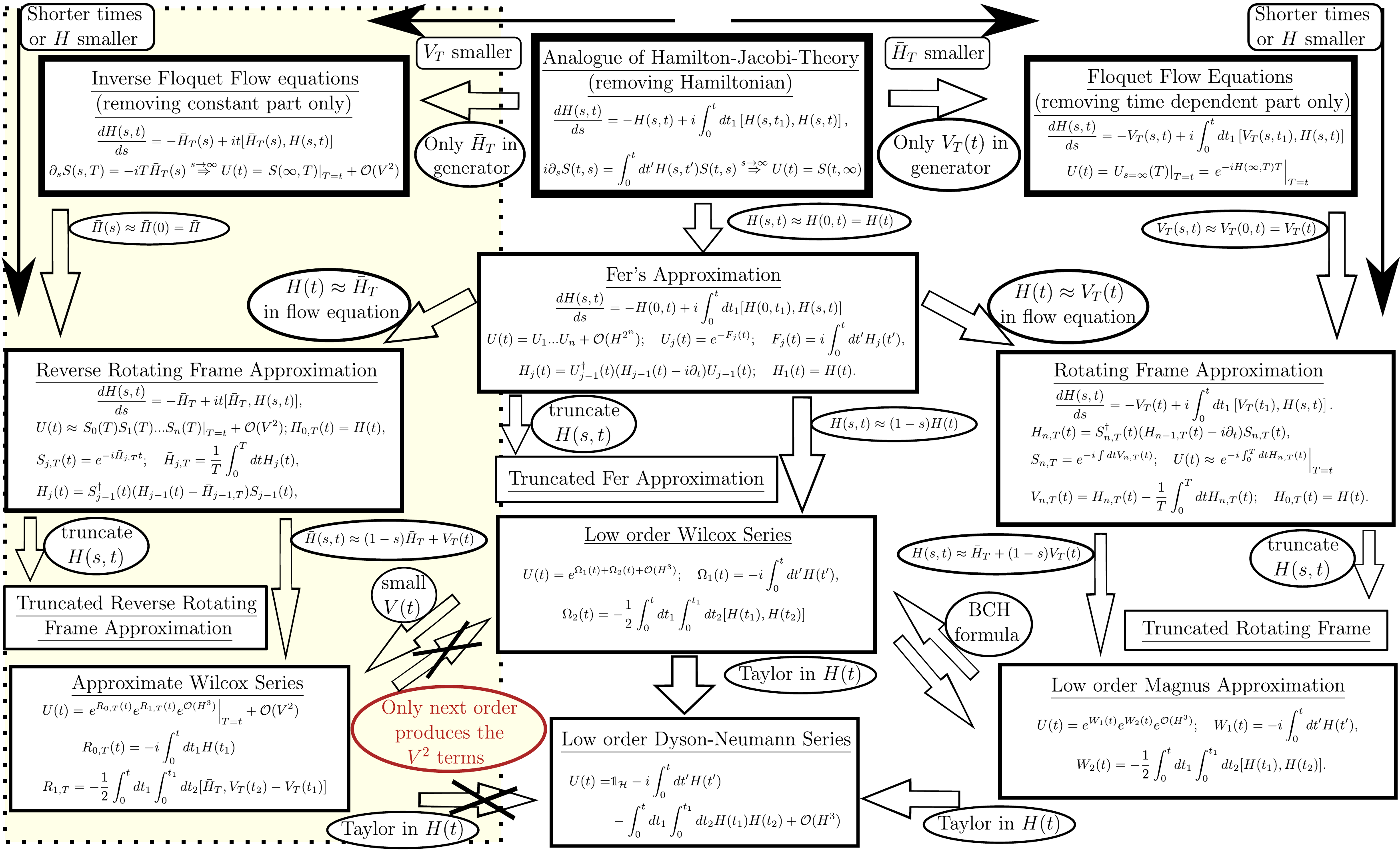}
	\caption{This illustration shows the relation between the different approximations discussed in the text. A new approximation that is implied by symmetry is shown in the dashed box on the left side of the figure. We supply these in later sections of the manuscript.  In the downward direction the approximations become progressively worse. To the left the approximations are expected to work better for larger constant parts of the Hamiltonian $\bar H$, and on the right for larger time-dependent parts of the Hamiltonian $V(t)$. The crossed out arrows signify that the result cannot be recovered without going to higher order in the approximate Wilcox series.}
	\label{fig:diagram1}
\end{figure*}

We are now in a position to draw a diagram in Fig.\ref{fig:diagram1} that relates the different approximations we discussed.  One may see by the symmetry of the diagram  in Fig.\ref{fig:diagram1} that there are some approximations still missing, which we marked in a dashed box. They will be the topic of the next section. 

\section{Large constant part in the Hamiltonian and the ``reverse" rotating frame approximation}
\label{sec:reverse_rot}
We would like to find a different flow equation to complete the rest of the diagram in the left-hand side of Fig.\ref{fig:diagram1}. This time we let $\bar H$ dominate in the generator, so that $\Sigma=-i\bar H_T(s)t$ and we find flow equations,
\begin{equation}
	\frac{dH(s,t)}{ds}=-\bar H_T(s)+it[\bar H_T(s),H(s,t)].
	\label{reverse_floquet_flow}
\end{equation}

Here, Eq.\eqref{reverse_floquet_flow} is not to be interpreted as an approximation, but rather it is constructed such that it removes all constant terms in the Hamiltonian. This is, 
$H(\infty,t)=V_T(\infty,t)$. More precisely, the time evolution operator in the transformed frame (at $s=\infty$) is $T\{e^{\int dt V(\infty,T)}\}=\mathbb{1}_{{H}}+\mathcal{O}(V^2)$ because $\int_0^T dt V_T(s,t)=0$. Therefore, to order $\mathcal{O}(V^2)$ we can neglect the contribution of the time evolution operator in the transformed frame. After setting $T=t$ (since $T$ was arbitrarily chosen) the time evolution is given as
\begin{equation}
	U(t)=\left.S(\infty,T)\right|_{T=t}+\mathcal{O}(V^2),
\end{equation}
with 
\begin{equation}
	\partial_s S(s,T)=-iT \bar H_T(s).
\end{equation}

Unlike the previous methods, Eq.\eqref{exact_flow} and Eq.\eqref{exact_flow2}, no integral appears on the right hand side of Eq.\eqref{reverse_floquet_flow}, which means that the coupling constants fulfill simple differential equations (and not complicated functional equations). However, we did not eliminate the need to calculate a time ordered exponential. Therefore, in its current form the method is not ideal.

Let us make the same approximation we made in both previous cases. If we assume that we may get rid of $\bar H$ swift enough that $s\approx 0$, then we may set $\bar H(s)\approx\bar H(0)=\bar H$ in the generator $\Sigma$ and the flow equations simplify as,
\begin{equation}
	\frac{dH(s,t)}{ds}=-\bar H_T+it[\bar H_T,H(s,t)],
	\label{approx_rrrot}
\end{equation}
where one lets $s$ run from $0$ to $1$ to remove $\bar H$ to lowest order in $t$. That is, we are now implementing a unitary transformation $S(t)=e^{-it\bar H}$. Because this (in the sense that we reduce out the constant part of the Hamiltonian) does the reverse of a rotating frame transformation we dub this the ``reverse" rotating frame approximation. Note that while we could calculate the unitary transformations exactly if $\bar H$ is not too complicated,  our formulation has the advantage that it enables us to use a truncated ansatz if needed. 

Going back to solve Eq.\eqref{approx_rrrot} exactly, we recognize that one may concatenate the reverse rotating frame transformations, which we will denote by $S_i$. The time evolution operator at times $T$ can then be approximated by a product of these $S_i$,
\begin{equation}
\begin{aligned}
 &U(t)\approx \left.S_0(T)S_1(T)... S_n(T)\right|_{T=t}+\mathcal{O}(V^2); H_{0,T}(t)=H(t),\\
 &S_{j,T}(t)=e^{-i\bar H_{j,T} t};\quad \bar H_{j,T}=\frac{1}{T}\int_0^{T}dt H_j(t),\\
 &H_j(t)=S_{j-1}^\dag(t)( H_{j-1}(t)-\bar H_{j-1,T})S_{j-1}(t),
\end{aligned}
\label{newexp}
\end{equation}
where again we were able to evaluate $U$ at $T=t$ since $T$ can be chosen arbitrarily.  The approximation in Eq.\eqref{newexp} is expected to work well in the limit of $\bar H_T\gg V_T$.

As in the previous cases let us make yet another approximation that follows the same structure as before. Namely, we first solve Eq.\eqref{approx_rrrot} for the case that we can neglect the commutator and find,
\begin{equation}
	H(s,t)\approx (1-s)\bar H_T+V_T(t).
\end{equation}
Reinserting this result one finds that,
\begin{equation}
	H(1,t)\approx V_T(t)+i\frac{t}{2}[\bar H_T,V_T(t)].
\end{equation}
Therefore, we may approximate by doing a time average
\begin{equation}
\begin{aligned}
\bar H_1&\approx \frac{i}{2T}\int_0^T dt\; t[\bar H_T,V_T(t)]\\
&=-\frac{i}{2T}\int_0^T dt_1\int_0^{t_1} dt_2[\bar H_T,V_T(t_2)-V_T(t_1)],
\end{aligned}
\end{equation}
where $\int_0^T dt V_T(t)=0$ and the Cauchy formula for repeated integration was used.
One may use this result in Eq.\eqref{newexp} to get an approximation to the time evolution operator. The result Eq.\eqref{newexp} reproduces the Wilcox series in the limit of small $V_T(t)$ to order $V_T(t)$.

Now that all approximations are in place, we have completed the approximation diagram in Fig.\ref{fig:diagram1}.  We now demonstrate the accuracy of these methods in various limits on a model system.

\section{Driven Ising model}
\label{sec:Ising}
To illustrate the quality of the approximations developed in the previous section and presented in Fig.\ref{fig:diagram1}, we will apply them to a one-dimensional spin chain and compare them with exact diagonalization results. We will consider the driven Ising model,
\begin{equation}
\begin{aligned}
&H(t)=\sum_i (J_z \sigma_i^z\sigma_{i+1}^z+B_z(t) \sigma_i^z+B_x(t)\sigma_i^x),\\
&B_x(t)=B\sin(\omega t);\quad B_z(t)=B\cos(\omega t),
\end{aligned}
\label{model}
\end{equation}
where $[\sigma_i^{x,y,z},\sigma_j^{x,y,z}]=0$ for $i\neq j$ and on-site they fulfill the Pauli algebra for spin-1/2 particles.

This model was chosen because it has much of the structure present in more complicated time-dependent problems because $[V(t_1),V(t_2)]\neq0$ in general, which is a common feature of many systems of interest.  Below we will derive expressions for all the different approximations (shown in Fig.\ref{fig:diagram1}) that are valid at times $T=\frac{2\pi}{\omega}$.

\subsection{Dyson-Neumann series}
Inserting our Hamiltonian, Eq.\eqref{model}, in Eq.\eqref{dysNeumann} we find that the following definitions are useful,
\begin{equation}
	\begin{aligned}
	&\bar H=\sum_i J_z \sigma_i^z\sigma_{i+1}^z,\\
	&V_y=B^2\sigma_i^y,\\
	&V_{yz}=2J_z B\sum_i \sigma_i^y(\sigma_{i+1}^z+\sigma_{i-1}^z),
	\end{aligned}
	\label{someHamiltons}
\end{equation}
and the time evolution operator is approximately given as,
\begin{equation}
	U(T)\approx \mathbb{1}-iT\bar H-\frac{T^2}{2}\left(\bar H^2+\frac{i}{\pi}V_{yz}-\frac{i}{\pi}V_y\right).
\end{equation}
One should note that in this case one was able to fully write down an analytical result.

\subsection{Magnus series and Wilcox series}
If we use Eq.\eqref{Magnus} and Eq.\eqref{Wilcox} we find that at $t=T$,
\begin{equation}
\hspace{-0.1cm}
	\begin{aligned}
	&\Omega_1=W_1=-iTJ_z\sum_i  \sigma_i^z\sigma_{i+1}^z,\\
	&\Omega_2=W_2=i\frac{T^2B}{\pi}\sum_i\sigma_i^y\left[\frac{B}{2}-J_z(\sigma_{i+1}^z+\sigma^z_{i-1})\right].\\
	\end{aligned}
\end{equation}
The approximate time evolution operators in the Magnus expansion $U_M$ and in the Wilcox approximation $U_W$ are,
\begin{equation}
	\begin{aligned}
	U_M(T)=e^{\Omega_1+\Omega_2};\quad U_W(T)=e^{W_1}e^{W_2}.
	\end{aligned}
\end{equation}
One was able to find analytical expressions for the exponents of the different contributions to the time evolution operator.

\subsection{Rotating frame approximation}
Here we may use Eq.\eqref{Rotatingframeflow} to find $H(1,t)$. The procedure is as follows. We start with a Hamiltonian that has the form of the original Hamiltonian, Eq.\eqref{model}, but with arbitrary couplings.  We then insert the Hamiltonian in the flow equations and add newly generated couplings to the Hamiltonian. This could be stopped at some point but here, because of the relatively simple structure of the external drive $V(t)=B_z(t) \sigma_i^z+B_x(t)\sigma_i^x$, we are able to reach a point when no new couplings are generated. The couplings that contribute are found to be the nine $\{\sigma_i^{x,y,z},\sigma_i^{x,y,z}\sigma_{i+1}^{x,y,z},\sigma_i^x\sigma_{i\pm1}^{y,z},\sigma_i^y\sigma_{i\pm1}^z\}$. To be more precise, the Hamiltonian in the rotating frame has the form,
\begin{equation}
\begin{aligned}
H_R&=\sum_i[C_x\sigma_i^x+C_y\sigma_i^y+C_z\sigma_i^z+C_{xx}\sigma_i^x\sigma_{i+1}^x\\
&+C_{xy}(\sigma_i^x\sigma_{i+1}^y+\sigma_i^x\sigma_{i-1}^y)+C_{yy}\sigma_i^y\sigma_{i+1}^y\\
&+C_{xz}(\sigma_i^x\sigma_{i+1}^z+\sigma_i^x\sigma_{i-1}^z)+C_{zz}\sigma_i^z\sigma_{i+1}^z\\
&+C_{yz}(\sigma_i^y\sigma_{i+1}^z+\sigma_i^y\sigma_{i-1}^z)]. \label{ansatzrotframeham}
\end{aligned}	
\end{equation}
The flow equations for the couplings, as well as the results, are given in Appendix \ref{appendix:flowrotf}. Averaging the results in Eq.\eqref{app:non-strobo-rotfrm}  over one period, we find that at stroboscopic times we have a Floquet Hamiltonian with couplings approximately given as,
\begin{equation}
\begin{aligned}
&C_x =C_{xy}=C_{xz}= 0,\\
&C_{xx}=\frac{3 J_z}{16}+\frac{3 J_z \omega ^2 J_2\left(\frac{4 B}{\omega }\right)}{8 B^2}-\frac{3 J_z \omega ^2 J_2\left(\frac{8 B}{\omega }\right)}{128 B^2}\\
&\hspace{0.7cm}+\frac{J_z \omega  J_1\left(\frac{8 B}{\omega }\right)}{16 B}-\frac{J_z \omega  J_1\left(\frac{4 B}{\omega }\right)}{2 B},\\
&C_{yy}=\frac{J_z}{4}+\frac{1}{2} J_z J_2\left(\frac{8 B}{\omega }\right)-\frac{J_z \omega  J_1\left(\frac{8 B}{\omega }\right)}{16 B},\\
&C_{yz}=\frac{1}{2} J_z J_1\left(\frac{8 B}{\omega }\right)+\frac{J_z \omega  J_2\left(\frac{4 B}{\omega }\right)}{4 B}-\frac{J_z \omega  J_2\left(\frac{8 B}{\omega }\right)}{16 B},\\
&C_{zz}=\frac{9 J_z}{16}-\frac{1}{2} J_z J_2\left(\frac{8 B}{\omega }\right)+\frac{3 J_z \omega ^2 J_2\left(\frac{8 B}{\omega }\right)}{128 B^2}\\
&\hspace{0.7cm}-\frac{3 J_z \omega ^2 J_2\left(\frac{4 B}{\omega }\right)}{8 B^2}+\frac{J_z \omega  J_1\left(\frac{4 B}{\omega }\right)}{2 B},\\
&C_{y}=\frac{\omega }{4}-\frac{1}{4} \omega  J_0\left(\frac{4 B}{\omega }\right)-B J_1\left(\frac{4 B}{\omega }\right),\\
&C_z=B J_2\left(\frac{4 B}{\omega }\right),
\end{aligned}
\end{equation}
where $J_n$ is the $n$-th Bessel function of the first kind.

The time evolution operator in this case is just,
\begin{equation}
	U(T)\approx e^{-iH_RT},
\end{equation}
with the couplings above. Similar to the Magnus case, we found an analytical expression for the exponent in the time evolution operator.

\subsection{Reverse rotating frame approximation}
Let us first find $H(1,t)$ according to Eq.\eqref{approx_rrrot}. The procedure is as follows. We start with a Hamiltonian that has the form of the original Hamiltonian, Eq.\eqref{model}, but with arbitrary couplings. We then insert the Hamiltonian in the flow equations and add newly generated couplings to the Hamiltonian. This could be stopped at some point but here we are able to reach a point when no new couplings are generated. The couplings that contribute are the five $\{\sigma_i^x,\sigma_i^z,\sigma_{i}^z\sigma_{i+1}^z,\sigma_{i}^y\sigma_{i\pm1}^z,\sigma_i^x\sigma_{i-1}^z\sigma_{i+1}^z\}$. Therefore, the Hamiltonian has the form,
\begin{equation}
\begin{aligned}
H_{RR}=&\sum_i C_x\sigma_i^x+C_z\sigma_i^z+C_{yz}(\sigma_i^y\sigma_{i+1}^z+\sigma_i^y\sigma_{i-1}^z)\\
&+C_{zz}\sigma_i^z\sigma_{i+1}^z+C_{xzz}\sigma_i^x\sigma_{i-1}^z\sigma_{i+1}^z.
\end{aligned}
\label{ansatzreverserotframeham}
\end{equation}

The flow equations that correspond to this are given in the Appendix \ref{app:reverserotframe}, and their solutions as well. After averaging the couplings Eq.\eqref{couplings_reverserotfrm} in the reverse rotating frame over one period, we find that the couplings are,
\begin{equation}
	\begin{aligned}
	& C_z=\bar C_{zz}=0;\quad  C_{yz}=\frac{B \omega ^2 \sin \left(\frac{8 \pi  J_z}{\omega }\right)}{4 \pi  \left(\omega ^2-16 J_z^2\right)},\\
	& C_x= C_{xzz}=\frac{B \omega ^2 \sin ^2\left(\frac{4 \pi  J_z}{\omega }\right)}{4 \pi  \left(\omega ^2-16 J_z^2\right)}.
	\end{aligned}.
		\label{UIsinginRRROTFrameCoupl}
\end{equation}

Therefore, the time evolution operator at stroboscopic times is approximately,
\begin{equation}
	U(T)\approx e^{-i\bar H T}e^{-i H_{RR}T},
	\label{UIsinginRRROTFrame}
\end{equation}
where $\bar H$ is given in Eq.\eqref{someHamiltons}. The first exponential factor in Eq. \eqref{UIsinginRRROTFrame}, $e^{-i\bar{H}T}$, should be interpreted as the transformation to the frame in which $H_{RR}$ and the couplings in Eq. \eqref{UIsinginRRROTFrameCoupl} are valid. Hence Eq. \eqref{UIsinginRRROTFrame} is valid only at $s=1$.

In addition we note that for this approximation we were able to give analytical expressions for the exponents in the time evolution operator.

\subsection{Fer approximation}
In the Fer approximation the flow equations, Eq.\eqref{flowingFer}, generate infinitely many terms. In its traditional form the Fer approximation would therefore not be applicable for such a system. Our method using flow equations, however, allows one to truncate those terms and include only terms that appeared in the rotating frame approximation and in the reverse rotating frame approximation. That is, we take 
\begin{equation}
\begin{aligned}
H_F&=\sum_i[C_x\sigma_i^x+C_y\sigma_i^y+C_z\sigma_i^z+C_{xx}\sigma_i^x\sigma_{i+1}^x\\
&+C_{xy}(\sigma_i^x\sigma_{i+1}^y+\sigma_i^x\sigma_{i-1}^y)+C_{yy}\sigma_i^y\sigma_{i+1}^y\\
&+C_{xz}(\sigma_i^x\sigma_{i+1}^z+\sigma_i^x\sigma_{i-1}^z)+C_{zz}\sigma_i^z\sigma_{i+1}^z\\
&+C_{yz}(\sigma_i^y\sigma_{i+1}^z+\sigma_i^y\sigma_{i-1}^z)+C_{xzz}\sigma_i^x\sigma_{i-1}^z\sigma_{i+1}^z].
\end{aligned}
\label{FeransatzHamiltonian}	
\end{equation}
The flow equations one finds for this ansatz are given in Appendix \ref{app:ferapprox}. While they are analytically accessible, the explicit expressions for the couplings are far too complicated to be illuminating.

\subsection{Truncated exact flow equations}

Since the Hamiltonian we consider in Eq.\eqref{model} has the form $H=H_0+e^{i\omega t}H_++e^{-i\omega t}H_-$, we may make use of Eq. \eqref{exactflowexample} to derive exact flow equations, which can be treated very conveniently numerically. Much like in the case of Fer's approximation this will generate infinitely many terms, which is why we took the same truncated ansatz, 
\begin{equation}
\begin{aligned}
H_{0,+,-}&=\sum_i[C_x\sigma_i^x+C_y\sigma_i^y+C_z\sigma_i^z+C_{xx}\sigma_i^x\sigma_{i+1}^x\\
&+C_{xy}(\sigma_i^x\sigma_{i+1}^y+\sigma_i^x\sigma_{i-1}^y)+C_{yy}\sigma_i^y\sigma_{i+1}^y\\
&+C_{xz}(\sigma_i^x\sigma_{i+1}^z+\sigma_i^x\sigma_{i-1}^z)+C_{zz}\sigma_i^z\sigma_{i+1}^z\\
&+C_{yz}(\sigma_i^y\sigma_{i+1}^z+\sigma_i^y\sigma_{i-1}^z)+C_{xzz}\sigma_i^x\sigma_{i-1}^z\sigma_{i+1}^z],
\end{aligned}	
\end{equation}
for all three parts of the Hamiltonian. The resulting flow equations are sufficiently opaque that we do not exhibit them.

The result from a numerical analysis is an effective Hamiltonian of the form,
\begin{equation}
\hspace{-0.3cm}
\begin{aligned}
H_{0}&=\sum_i[C_y\sigma_i^y+C_z\sigma_i^z+C_{xx}\sigma_i^x\sigma_{i+1}^x+C_{yy}\sigma_i^y\sigma_{i+1}^y\\
&\hspace{0.7cm}+C_{zz}\sigma_i^z\sigma_{i+1}^z+C_{yz}(\sigma_i^y\sigma_{i+1}^z+\sigma_i^y\sigma_{i-1}^z)].
\end{aligned}	
\end{equation}

Other terms from the truncated ansatz vanish up to numerical accuracy.  However, they appear during the flow. While this method does not offer us analytic expressions for the couplings it still has advantages over brute force exact diagonalization. One important advantage is that the method is scalable:  one may include as many terms in the ansatz as desired and therefore arrive at different levels of numerical costs. Such an ansatz may be motivated by physical considerations or mathematically by perturbation theory, such as we used. Furthermore, by using this method one has an explicit expression in terms of operators and may therefore do semi-analytical follow-up work.

\section{Comparing all the approximations}
\label{sec:allApprox}
In this section we compare the validity of the different approximations discussed in the previous sections by calculating the $l_2$ distance, 
\begin{equation}
	l_2(U_A,U_E)=\frac{1}{2\sqrt{D}}\sqrt{tr((U_A-U_E)(U_A-U_E)^\dag)},
\label{eq:l2_distance}
\end{equation}
between the various approximate time evolution operators, $U_A$, and the exact time evolution operator, $U_E$. The distance that is found via exact diagonalization of systems with up to 16 sites using the QuSpin package.\cite{SciPostPhys.2.1.003} All evolution operators are evaluated at stroboscopic times and the Hilbert space has dimension $D$. When both operators are unitary, $l_2(U_A,U_E) \in [0,1]$. Zero corresponds to perfect agreement ($U_A=U_E$), and unity corresponds to maximally separated unitary operators.

One should note that this measure sometimes overestimates errors. This may occur, for example,  when evaluating the time evolution of local observables. For instance \cite{2018arXiv180611123H,2018arXiv181205876S}, found that local observables for a Trotterized time evolution like Eq.\eqref{trotterized} are more robust to an increase in time-step size  $dt$ than previously thought. The reason for this is that local observables when evolved for sufficiently short times occupy only a small part of the total Hilbert space. The method above as an error estimate has contributions from all parts of the Hilbert space and therefore may overestimate the error for certain parts of the Hilbert space. We therefore like to think of it as a worst case estimate or as an estimate for global properties of the time evolution. While it is very interesting to find out how well these different methods work for different observables we will not attempt to discuss such behavior that is specific to a certain operator but rather discuss the generic properties captured in Eq.\eqref{eq:l2_distance}.

\begin{figure}
	\centering
	\includegraphics[width=1\linewidth]{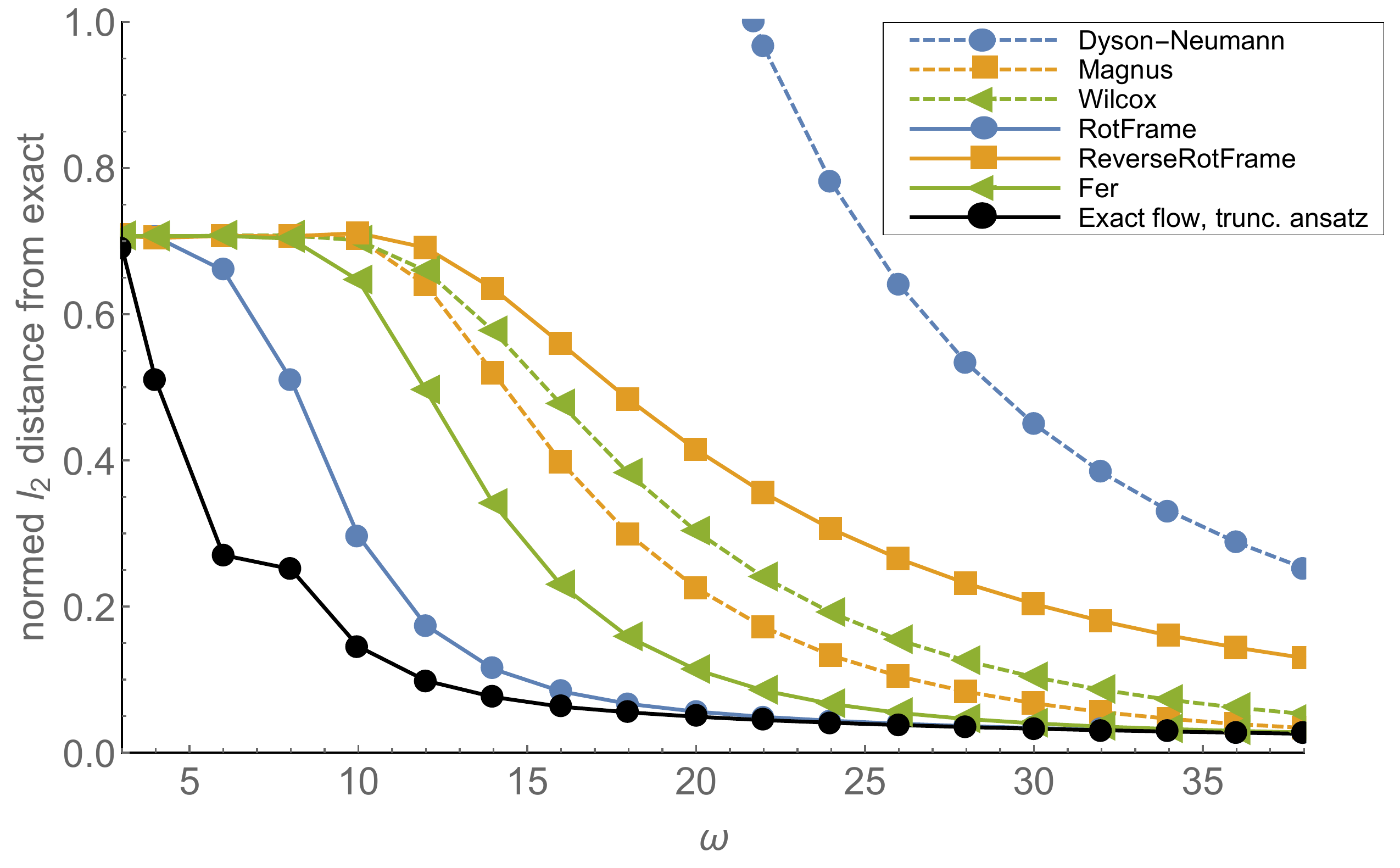}
	\caption{Plot of the $l_2$ distance, Eq.\eqref{eq:l2_distance}, between the different approximations and the exact time evolution operator for strong driving $B=4$, $J_z=1$ and $L=14$ sites. Evidently, all approximations perform better as the large frequency limit is approached. }
	\label{fig:largedrivingomega}
\end{figure}

We wish to determine how well the different approximations perform as a function of frequency for different strengths of couplings. Let us first look at the limit of large driving strength, $B$. From Fig.\ref{fig:largedrivingomega}, one may see that, as expected, the Dyson-Neumann approximation (dashed blue or in print dashed gray with circle markers) has the worst performance, and even reaches values above $1$, because it is not unitary. The reverse rotating frame approximation (solid orange or in print solid gray with rectangle markers) performs poorly, too, which comes as no surprise because it neglects $B^2$ terms. We may choose the Magnus approximation over the Wilcox approximation because the Magnus approximation is more accurate. 

A recurring theme we find is that approximations that make life simpler generally perform more poorly:  multiple less complicated matrix exponentials in the Wilcox case would have been easier to calculate than one complicated matrix exponential in the Magnus case.  The Fer approximation performs slightly worse than the rotating frame approximation, which is likely due to the need to truncate it at an arbitrary point. From the analytically accessible approximations, the rotating frame approximation performs best.  However, even it is outperformed by the exact flow equations, including the case of a truncated ansatz.  This example demonstrates that the flow equations are indeed especially useful when looking for Floquet Hamiltonians.
 
\begin{figure}
 	\centering
 	\includegraphics[width=1\linewidth]{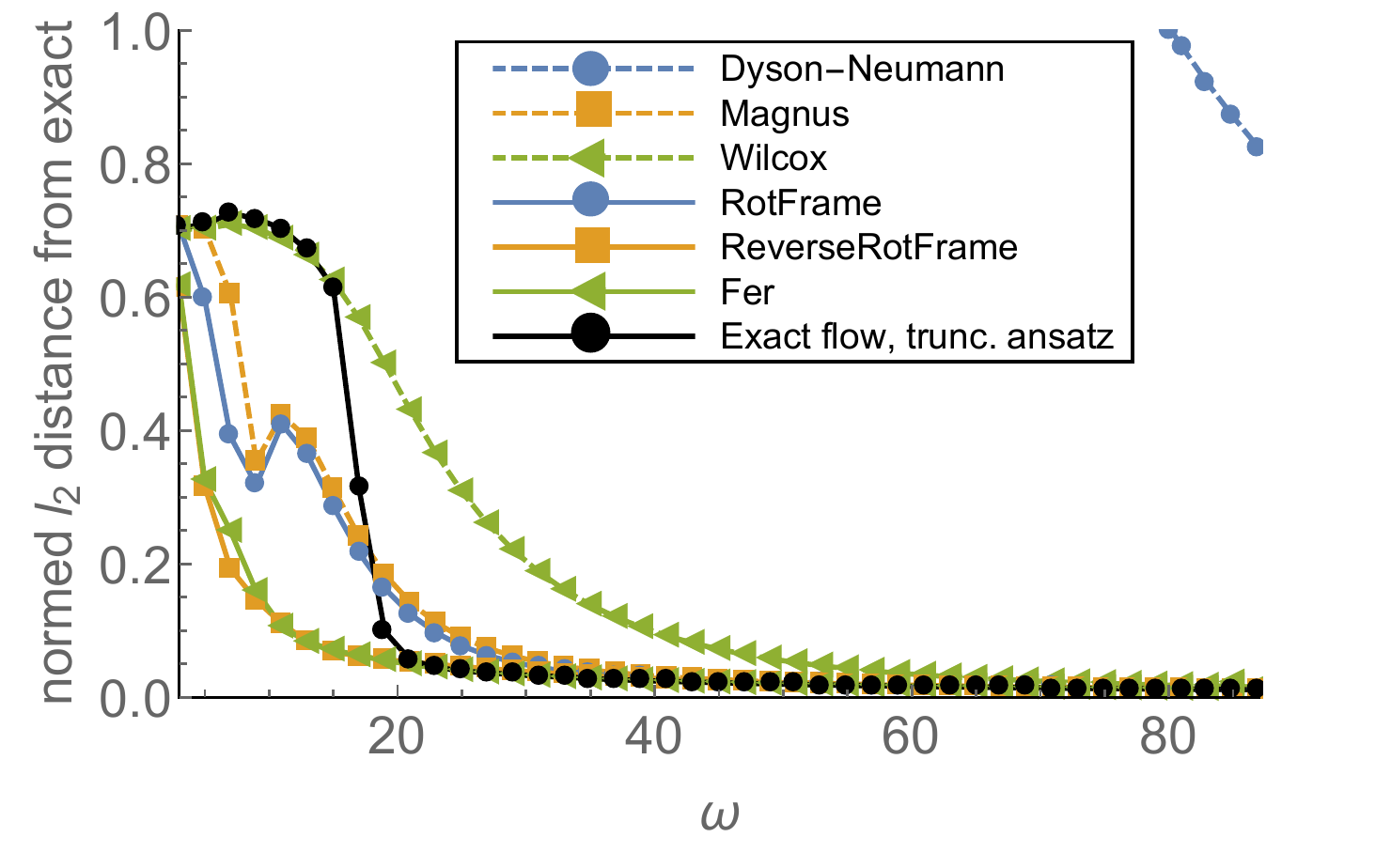}
 	\caption{Plot of the $l_2$ distance between the different approximations and the exact time evolution operator for weak driving $B=1$, and strong static interaction $J_z=4$, for a chain of $L=14$ sites.  }
 	\label{fig:largeconstgomega}
 \end{figure}

 Next we consider the case of strong static parts in the Hamiltonian. The plot is given in Fig.\ref{fig:largeconstgomega}. We find that for the Dyson-Neumann series, for almost the full range of values considered, unitarity is completely broken and $l_2$ does not even appear within the range $[0,1]$. The reverse rotating frame approximation does best for these large couplings.  For most of the range of values the Fer approximation performs similarly. As we found with a strong drive, the Magnus approximation outperforms the Wilcox approximation. The rotating frame approximation is only a slight improvement over the Magnus approximation. The result from the truncated but exact flow equations for the range $\omega\gtrsim20$ is comparable to the best approximations. For the range of values $\omega\lesssim20$---presumably because of the truncation scheme---it becomes uncontrolled. It is worth mentioning that, for the truncated but exact flow equations, fewer couplings contribute to the effective Hamiltonian than in the Fer case.  Some of the couplings that appear in the Fer case are zero for truncated flow equations, which to some extent explains the shorter range of validity of the approximation.
 
\begin{figure*}
	\centering
	\includegraphics[width=1\linewidth]{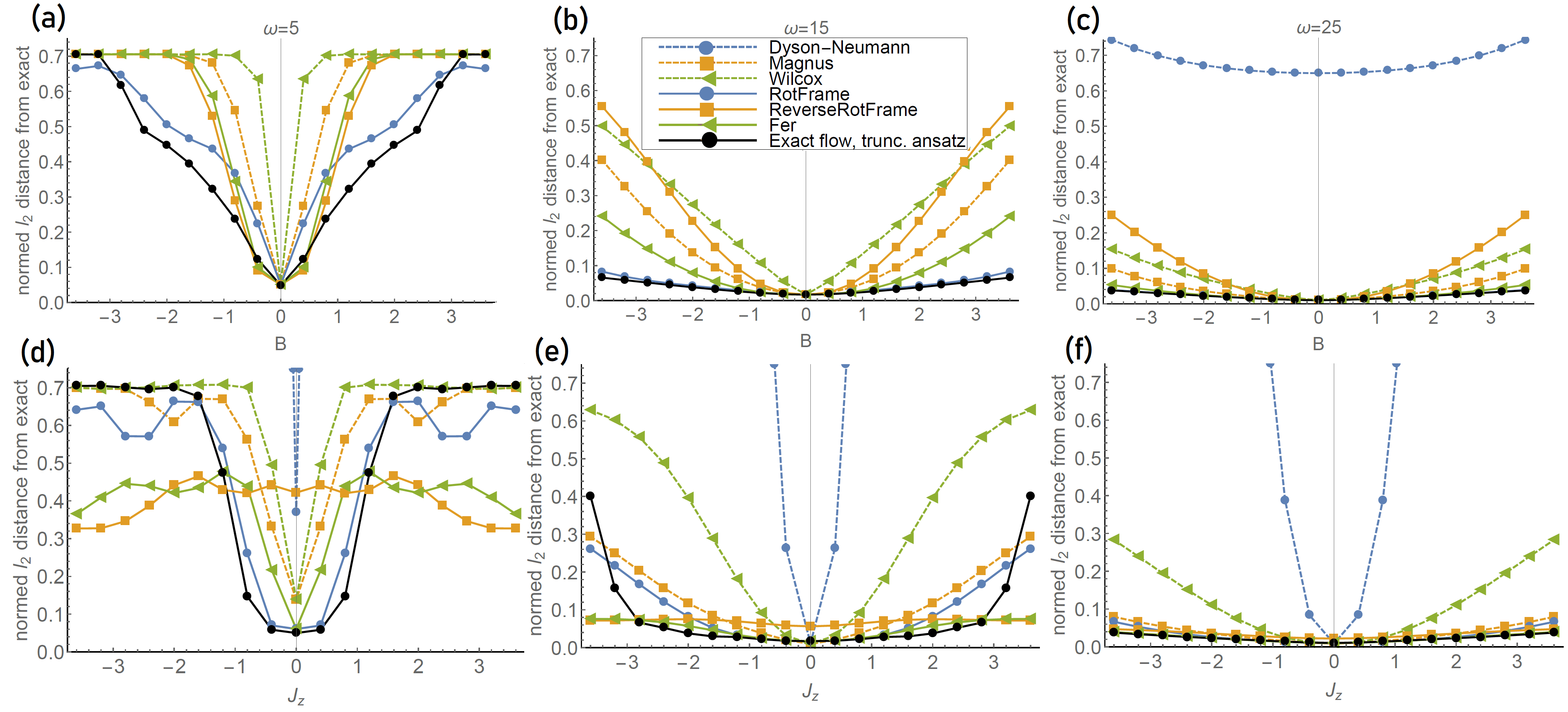}
	\caption{Plot of the $l_2$ distance between the different approximations and the exact time evolution operator as a function of the couplings for a chain with $L=16$ sites. The upper row has $J_z=1$ fixed with $B$ being varied. The lower row has $B=1$ fixed and $J_z$ is varied. For both cases the left column has $\omega=5$, the middle column $\omega=15$ and the right column $\omega=25$. }
	\label{fig:plotscouplings}
\end{figure*}
Let us next look at how the approximations behave as functions of the couplings. We see this in Fig.\ref{fig:plotscouplings} for three different frequencies. As expected we find the Dyson-Neumann series performs the worst across the board, and the Wilcox approximation is second worst in most cases. The Magnus approximation, as expected, is typically third worst. For increased magnetic driving we see that the exact but truncated flow equations and the rotating frame approximation are the most reliable with the results for the truncated flow equations being slightly better. 

In general we can see that the exact but truncated flow equations for a wide range of variables (yet not all) yield the most reliable results. The reason it does not always perform better is the arbitrarily chosen truncation point. We find that the ``reverse" rotating frame approximation we introduce is most useful in the intermediate frequency regime at comparatively strong constant parts in the Hamiltonian. To make the results more accessible we provide Table \ref{tableWhenTo} summarizing the results for the \emph{first} iteration of each procedure.

\begin{table*}
	\begin{tabular}{|l||c|c|c||c|c|c||c|c|c||c|c|}\hline
		&\multicolumn{3}{c}{$t^{-1}\sim\omega\ll H_0,V$}&  \multicolumn{3}{||c|}{$t^{-1}\sim\omega\sim H_0,V$}&  \multicolumn{3}{||c||}{$t^{-1}\sim\omega \gg H_0,V$}&$H_F$&Remarks  \\\hline
		
		&any&  $V\gg H_0$&  $H_0\gg V$&any&  $V\gg H_0$&  $H_0\gg V$&any&  $V\gg H_0$&  $H_0\gg V$& &   \\\hline\hline
		\makecell{Time ordered exponential}&  \yes &  \yes&\yes&  \yes&\yes&  \yes&\yes&  \yes&  \yes& \no& \\\hline
		\makecell{Exact flow equations\\(removing $H$ or $H_0$)}&  \yes &  \yes&\yes&  \yes&\yes&  \yes&\yes&  \yes&  \yes& \no&\makecell{Inaccessible analytically,\\
		numerics harder than\\ time ordered exponential } \\\hline
	    \makecell{Exact flow equations\\ (removing $V$)}&  \yes&  \yes&\yes&  \yes&\yes&  \yes&\yes&  \yes&  \yes&\yes &\makecell{Inaccessible analytically,\\ numerics comparable to\\ time ordered exponential}\\\hline
	    \makecell{Truncated exact flow\\(removing $V$)}&  \no&  \no&\no&  \yes&\yes&  \yes&\yes&  \yes&  \yes&\yes &\makecell{Inaccessible analytically,\\ numerics require a few\\ flow equations and one\\ matrix exponential}\\\hline
	    Fer expansion&  \no&  \no&\no&  \yes&\yes&  \yes&\yes&  \yes&  \yes&\no  &\makecell{If truncated exponents\\ accessible analytically,\\ numerics require easier\\ flow equations but two\\ matrix exponentials}\\\hline
	    Rotating frame&  \no&  \no&\no&  \no&\yes&  \no&\yes&  \yes&  \yes&\yes &\makecell{$H_F$ accessible by analytical\\ treatments sometimes\\ without truncation,\\ numerics require easier\\ flow equations and one\\ matrix exponential}\\\hline
	    Reverse rot. frame&  \no&  \no&\no&  \no&\no&  \yes&\no&  \no&  \yes&\no &\makecell{Exponents accessible \\analytically sometimes\\ without truncation,\\ numerics require easier\\ flow equations but two\\ matrix exponentials}\\\hline
	    Magnus approximation&  \no&  \no&\no&  \no&\no&  \no&\yes&  \yes&  \yes&\yes &\makecell{$H_F$ accessible analytically,\\ numerics require  one\\ matrix exponential}\\\hline
	    Wilcox approximation&  \no&  \no&\no&  \no&\no&  \no&\yes&  \yes&  \yes&\no &\makecell{Exponents accessible\\ by analytical means,\\ numerics require two\\ matrix exponential}\\\hline
	    Dyson-Neumann&  \no&  \no&\no&  \no&\no&  \no&\yes&  \yes&  \yes&\no &Analytically accessible\\\hline
	\end{tabular}
\caption{Summary of performance quality of various approximations discussed in the text.}
\label{tableWhenTo}
\end{table*}

It should be emphasized that the checkmarks in the table do not capture that the Magnus approximation is vastly better than the Wilcox approximation or the Dyson-Neumann approximation, but it should serve as a qualitative guide on which method to use. We would also like to stress that the reverse rotating frame approximation makes a regime easier to access analytically when it is not covered by the other approximations that are analytically tractable. One should also note that it can very easily be combined with a first order Magnus approximation, which would turn the two red checkmarks in the high frequency or short time regime green because this reintroduces the order $V^2$ terms that were neglected.

\section{Conclusion}
\label{sec:conclusions}
In this work, we introduce an analogue of Hamilton-Jacobi theory for the time-evolution operator of a quantum many-particle system. The theory offers a useful approach to develop approximations to the time-evolution operator, and also provides a unified framework and starting point for many well-known approximations to the time-evolution operator.  

In the process we found a novel approximation to the time-evolution operator, which is accurate if the constant part of the Hamiltonian is large compared to the time-dependent part. This approximation may be useful in cavity QED applications as discussed earlier or more generally in cases where the constant part of Hamiltonian is large enough that the Magnus expansion will be an insufficient approximation despite a small external driving strength compared to the driving frequency. We were also able to show that one set of flow equations we derived in a prior work turns out to be especially powerful since it offers the best approximation, even when truncated, to the time-evolution operator while still being numerically easily accessible. Unlike time ordered exponentials, however, it also facilitates easy access to the Floquet Hamiltonian since coefficients in the Floquet Hamiltonian can be calculated directly, which opens the road to semi-analytic discussions of systems that are otherwise inaccessible. We hope that the flowchart we provided  in Fig.\ref{fig:diagram1} will guide an understanding of the connections between different popular approximations.  In addition, we hope Table\ref{tableWhenTo} we provided will make it easy for a reader to appropriately choose the right approximation for any problem encountered.

\acknowledgments
We are grateful for funding under NSF DMREF Grant no DMR-1729588 and NSF Materials Research Science and Engineering Center Grant No. DMR-1720595.  During the writing of the manuscript, P.L. was supported by the Scientific Discovery through Advanced Computing (SciDAC) program funded by the US Department of Energy, Office of Science, Advanced Scientific Computing Research and Basic Energy Sciences, Division of Materials Sciences and Engineering. GAF acknowledges support from a Simons Fellowship, and a 
QuantEmX grant from ICAM and the Gordon and Betty Moore Foundation through Grant GBMF5305.

\bibliographystyle{unsrt}
\bibliography{literature}

\newpage
\appendix
\begin{widetext}
\section{Flow equations for rotating frame}
\label{appendix:flowrotf}
We find that the flow equations for the rotating frame transformation of Eq.~\eqref{ansatzrotframeham} for the Ising model are 
	\begin{equation}
	\begin{aligned}
	&\frac{dC_x(s)}{ds}=2 C_y(s) F_2(t)-B\sin(\omega t);&\frac{dC_y(s)}{ds}=2 C_z(s) F_1(t)-2 C_x(s) F_2(t),\\
	&\frac{dC_z(s)}{ds}=-2 C_y(s) F_1(t)-B\cos(\omega t);&\frac{dC_{xx}(s)}{ds}=4 C_{xy}(s) F_2(t),\\
	&\frac{dC_{xy}(s)}{ds}=-2 C_{xx}(s) F_2(t)+2 C_{xz}(s) F_1(t)+2 C_{yy}(s) F_2(t);&\frac{dC_{yy}(s)}{ds}=4 C_{yz}(s) F_1(t)-4 C_{xy}(s) F_2(t),\\
	&\frac{dC_{xz}(s)}{ds}=2 C_{yz}(s) F_2(t)-2 C_{xy}(s) F_1(t);&\frac{dC_{zz}(s)}{ds}=-4 C_{yz}(s) F_1(t),\\
	&\frac{dC_{yz}(s)}{ds}=-2 C_{xz}(s) F_2(t)-2 C_{yy}(s) F_1(t)+2 C_{zz}(s) F_1(t),&
	\end{aligned}
	\end{equation}
where $F_1(t)=\frac{B-B\cos (\omega t )}{\omega }$ and $F_2(t)=\frac{B\sin (\omega t )}{\omega }$
and with the initial conditions
\begin{equation}
	\begin{aligned}
	&C_{xx}(0)=C_{xy}(0)=C_{yy}(0)=0,\\
	&C_{xz}(0)=C_{yz}(0)=C_y(0)=0,\\
	&C_{zz}(0)=J_z;\quad C_x(0)=B\sin(\omega t),\\ 
	&C_z(0)=B\cos(\omega t).
	\end{aligned}
\end{equation}

Solving these equations one finds
\begin{equation}
	\begin{aligned}
	&C_{xx}(1,t)= J_z \sin ^2(\omega t ) \sin ^4\left(\frac{2 B \sin \left(\frac{\omega t }{2}\right)}{\omega }\right),\\
	&C_{xy}(1,t)= J_z \sin \left(\frac{\omega t }{2}\right) \sin (\omega t ) \sin ^2\left(\frac{2 B \sin \left(\frac{\omega t }{2}\right)}{\omega }\right) \sin \left(\frac{4 B \sin \left(\frac{\omega t }{2}\right)}{\omega }\right),\\
	&C_{xz}(1,t)= \frac{1}{4} J_z \left(2 \sin (2 \omega t ) \sin ^4\left(\frac{2 B \sin \left(\frac{\omega t }{2}\right)}{\omega }\right)+\sin (\omega t ) \sin ^2\left(\frac{4 B \sin \left(\frac{\omega t }{2}\right)}{\omega }\right)\right),\\
	&C_{yy}(1,t)= J_z \sin ^2\left(\frac{\omega t }{2}\right) \sin ^2\left(\frac{4 B \sin \left(\frac{\omega t }{2}\right)}{\omega }\right),\\
	&C_{yz}(1,t)= \frac{1}{2} J_z \sin ^3\left(\frac{\omega t }{2}\right) \sin \left(\frac{8 B \sin \left(\frac{\omega t }{2}\right)}{\omega }\right)+J_z \sin \left(\frac{\omega t }{2}\right) \cos ^2\left(\frac{\omega t }{2}\right) \sin \left(\frac{4 B \sin \left(\frac{\omega t }{2}\right)}{\omega }\right),\\
	&C_{zz}(1,t)= \frac{1}{16} J_z \left(8 \sin ^4\left(\frac{\omega t }{2}\right) \cos \left(\frac{8 B \sin \left(\frac{\omega t }{2}\right)}{\omega }\right)+8 \sin ^2(\omega t ) \cos \left(\frac{4 B \sin \left(\frac{\omega t }{2}\right)}{\omega }\right)+4 \cos (\omega t )+3 \cos (2 \omega t )+9\right),\\
	&C_{x}(1,t)= \frac{1}{4} \left(2 B \sin (\omega t ) \cos \left(\frac{4 B \sin \left(\frac{\omega t }{2}\right)}{\omega }\right)-\omega  \cos \left(\frac{\omega t }{2}\right) \sin \left(\frac{4 B \sin \left(\frac{\omega t }{2}\right)}{\omega }\right)\right),\\
	&C_{y}(1,t)= \frac{1}{2} \omega  \sin ^2\left(\frac{2 B \sin \left(\frac{\omega t }{2}\right)}{\omega }\right)-B \sin \left(\frac{\omega t }{2}\right) \sin \left(\frac{4 B \sin \left(\frac{\omega t }{2}\right)}{\omega }\right),\\
	&C_{z}(1,t)= \frac{1}{4} \left(\omega  \sin \left(\frac{\omega t }{2}\right) \sin \left(\frac{4 B \sin \left(\frac{\omega t }{2}\right)}{\omega }\right)+2 B (\cos (\omega t )-1) \cos \left(\frac{4 B \sin \left(\frac{\omega t }{2}\right)}{\omega }\right)\right).
	\end{aligned}
	\label{app:non-strobo-rotfrm}
\end{equation}
\newpage
\end{widetext}

\section{Reverse rotating frame flow equations}
\label{app:reverserotframe}
The flow equations that are found with the Hamiltonian given by \eqref{ansatzreverserotframeham} when inserted in \eqref{approx_rrrot} are
\begin{equation}
	\begin{aligned}
	&\frac{dC_{yz}(s)}{ds}=-2 J_z t C_x(s)-2 J_z t C_{xzz}(s),\\
	&\frac{dC_{zz}(s)}{ds}=-J_z;\quad \frac{dC_z(s)}{ds}=0,\\
	&\frac{dC_{xzz}(s)}{ds}=\frac{dC_x(s)}{ds}=4 J_z t C_{yz}(s),
	\end{aligned}
\end{equation}
with initial conditions
\begin{equation}
	\begin{aligned}
	&C_{yz}(0)=C_{xzz}(0)=0;\quad C_{zz}(0)=J_z,\\
	& C_x(0)=B \sin (\omega t );\quad C_z(0)=B \cos (\omega t ).\\
	\end{aligned}
\end{equation}
Solving this we find that
\begin{equation}
\begin{aligned}
&C_{zz}(1,t)=0;\quad C_z(1,t)=B \cos (\omega t ),\\
&C_x(1,t)=B \cos ^2(2 J_z t) \sin (\omega t ),\\
&C_{xzz}(1,t)=-B \sin ^2(2 J_z t) \sin (\omega t ),\\
&C_{yz}(1,t)=B \cos ^2(2 J_z t) \sin (\omega t ).
\end{aligned}
\label{couplings_reverserotfrm}
\end{equation}

\section{Fer approximation flow equations}
\label{app:ferapprox}
The flow equations that are found with the Hamiltonian given by \eqref{FeransatzHamiltonian} when inserted in \eqref{flowingFer} and are found as
\begin{equation}
	\begin{aligned}
	&\frac{d C_{xx}(s)}{ds}=4 C_{xy}(s) f_z(t),\\
	&\frac{d C_{xy}(s)}{ds}=-2 C_{xx}(s) f_z(t)+2 C_{xz}(s) f_x(t)+2 C_{yy}(s) f_z(t),\\
	&\frac{d C_{yy}(s)}{ds}=4 C_{yz}(s) f_x(t)-4 C_{xy}(s) f_z(t),\\
	&\frac{d C_{xz}(s)}{ds}=-2 C_{xy}(s) f_x(t)+2 J_z t C_y(s)+2 C_{yz}(s) f_z(t),\\
	&\frac{d C_{yz}(s)}{ds}=-2 J_z t C_x(s)-2 C_{xz}(s) f_z(t)-2 J_z t C_{xzz}(s)\\
		&\quad-2 C_{yy}(s) f_x(t)+2 C_{zz}(s) f_x(t),\\
	&\frac{d C_{zz}(s)}{ds}=-4 C_{yz}(s) f_x(t)-J_z,\\
	&\frac{d C_{xzz}(s)}{ds}=4 J_z t C_{yz}(s),\\
	&\frac{d C_x(s)}{ds}=-B \sin (t \omega )+2 C_y(s) f_z(t)+4 J_z t C_{yz}(s),\\
	&\frac{d C_y(s)}{ds}=-2 C_x(s) f_z(t)-4 J_z t C_{xz}(s)+2 C_z(s) f_x(t),\\
	&\frac{d C_z(s)}{ds}=-B \cos (t \omega )-2 C_y(s) f_x(t),
	\end{aligned}
\end{equation}
where $f_x(t)=\frac{B (1-1 \cos (t \omega ))}{\omega }$ and $f_z(t)=\frac{B \sin (t \omega )}{\omega }$.

The initial conditions are
\begin{equation}
	\begin{aligned}
	&C_{xx}(0)=C_{xy}(0)=C_{yy}(0)=C_{xz}(0),\\
	&\hspace*{1cm}=C_{yz}(0)=C_{xzz}(0)=C_y(0)=0,\\
	&C_x(0)=B \sin (\omega t);\quad C_z(0)=B \cos (\omega t);\quad C_{zz}(0)=J_z.
	\end{aligned}
\end{equation}
The solutions are too tedious to write out and therefore not given.

\end{document}